\documentclass[twocolumn]{aastex701}
\usepackage{tabularx}
\usepackage{amsmath}
\usepackage{amssymb}
\usepackage{bm}
\usepackage{graphicx}
\usepackage[export]{adjustbox} 

\newcommand{\HI}{H{\small I} }
\newcommand{\diff}{\mathop{}\!\mathrm{d}} 

\graphicspath{{./}{figures/}}

\begin{document}

\title{Quantifying Foreground Contamination in the Dark Ages 21\,cm Power Spectrum Using an Electromagnetically Simulated Dipole Antenna Atop a Dielectric Surface}

\author{Willow Smith}
\affiliation{Department of Physics, Brown University, Providence, RI, USA}
\email{}

\author{Jonathan C. Pober}
\affiliation{Department of Physics, Brown University, Providence, RI, USA}
\email{}

\author{Bang D. Nhan}
\affiliation{National Radio Astronomy Observatory, Charlottesville, VA, USA}
\email{}

\author{Jack O. Burns}
\affiliation{Center of Astrophysics and Space Astronomy, University of Colorado Boulder, Boulder, CO, USA}
\email{}

\author{Ronald S. Polidan}
\affiliation{Lunar Resources, Inc., Houston, TX, USA}
\email{}

\begin{abstract}

The highly-redshifted 21\,cm signal from the cosmic Dark Ages presents an exciting frontier for cosmology, with the potential to observe a large number of Fourier modes of the cosmic density field in the absence of complicating astrophysical phenomena. Because the Earth's ionosphere and human-generated interference affect these low radio frequencies, the lunar far side --- the most radio-quiet region in the inner solar system --- is considered the ideal site to conduct such an experiment. The bright synchrotron foregrounds at these frequencies are expected to be spectrally smooth and thus occupy a subset of spectral Fourier modes, leaving an observable ``window'' to the Dark Ages signal. However, in practice, spectral structure arises in foregrounds due to instrumental artifacts, leading to a spillover of foreground power into the observable window. In this paper, we quantify this spillover for a set of simulated visibility measurements given a number of antenna beams, sky models, and baseline configurations. Notably, we examine the effects on foreground spillover due to variations in the lunar regoliths using two numerically simulated beam models with 1 and 4 layers of substrate materials underneath. We find that the regolith material properties can induce unwanted spectral structure in the beams, which potentially prevents foreground power suppression at the levels required. The detailed spectral behavior depends on the lunar regolith model used, making it paramount to produce both accurate models of foregrounds and of the lunar regolith for Dark Ages cosmology.

\end{abstract}

\section{Introduction} 
\label{sec:intro}

One of the next frontiers of groundbreaking observations is found in the cosmic Dark Ages, the period of time between the last scattering that created the Cosmic Microwave Background (CMB) and the formation of the first stars (Cosmic Dawn). During this period, the universe is absent of any complex astrophysics and mostly consists of neutral Hydrogen \HI and dark matter interacting gravitationally. The lack of stars or other light emitting sources gives the period its name, but the occasional CMB photon is absorbed and emitted from \HI through the hyperfine transition which corresponds to a 21\,cm wavelength (see \cite{Pritchard_2008} for a review). 

Observation of the 21\,cm signal would offer a direct probe of the density and distribution of \HI which in turns traces dark matter, typically quantified in the form of a power spectrum in Fourier space; because modes along the line of sight ($k_\parallel$) are measured via the redshifting of the 21\,cm signal (and hence come from the observed spectrum), while modes in the plane of the sky ($k_\perp$) are measured via mapping, experimentalists often work in terms of a 2D, or cylindrically averaged, power spectrum: $P(k_{\perp}, k_{\parallel})$. The 21\,cm power spectrum offers constraints on cosmological parameters (e.g., 
$\Omega_{M}$, $H_{0}$, etc.) as well as a direct probe of exotic physics models like charged dark matter \citep{Mu_oz_2018}. In fact, in the cosmic variance limit, the 21\,cm power spectrum offers one to three orders of magnitude improvement of on the constraints of cosmological parameters \citep{Mondal_2023}. The cosmic Dark Ages are therefore often regarded as a cosmological treasure trove of statistical measurements.

However, observing the 21\,cm signal is an open question with several challenges. The signal is redshifted to wavelengths that the Earth's ionosphere is often opaque to and that overlap significantly with several sources of radio frequency interference (RFI) \citep{Vedantham_2016}. Additionally, due to the rarity of the transition and subsequent faintness, a very large radio instrument is required for the necessary sensitivity to observe the signal. Therefore, current proposals aim to put a several square kilometer radio interferometer on the far side of the Moon --- where the Moon would shield from Earth's RFI and the lack of ionosphere would leave the signal intact \citep{burns2021global21cmcosmologyfarside}.

One such concept is \textit{FarView}, an in-situ manufactured 100,000 dipole radio interferometer \citep{polidan2024farviewinsitumanufacturedlunar, burns2026farview}. It has been shown for a radio interferometer on the scale of \textit{FarView}, that a 10$\sigma$ signal-to-noise ratio (SNR) can be obtained for the $z=30$ power spectrum in the absence of foregrounds \citep{smith2025detecting21cmsignal}.  Even in the case that partial subtraction can be achieved, a detection is still possible. However, subtracting foregrounds presents a massive challenge that no analysis method has yet reached the precision for for 21~cm cosmology.

Because of its spectral smoothness, one might expect foreground power in Fourier space to occupy only low $k_{\parallel}$ modes, which would make foreground subtraction less pertinent. However, spectral leakage in foreground modes causes foregrounds to spread out to higher $k_{\parallel}$ with increasing $k_{\perp}$ creating what's called the \textit{foreground wedge}. The detailed properties of this spectral leakage depend on the specific method of power spectrum estimation used, but the wedge feature itself is quite generic. We describe this process in Section \ref{subsec:spill} for a delay spectrum analysis which we employ in this paper, but for further detail on the foreground wedge see \cite{Bowman_2009}, \cite{Datta_2010}, \cite{Morales_2012}, and \cite{Parsons_2012b}.

The remaining modes, not dominated by the bright foreground wedge, offer a ``window'' to the 21\,cm signal. However, the extent of this ``window'' is dependent on a number of additional factors. Firstly, the choice of window or tapering function used in the frequency Fourier transform and the finiteness of bandwidth cause additional foreground power to spill over beyond the foreground wedge. Secondly, spectral structure introduced by the antenna's electromagnetic response, i.e., the antenna primary beam, can further complicate the signal. This spectral structure can be understood through the antenna's electrical coupling with other antennas, the ground (lunar regolith), and the sky. Recent analyses of the electrical coupling of a dipole antenna atop lunar regolith suggests that the regolith's dielectric properties play a complex role in spectral contamination \citep{hibbard2024medea, rotermund2026electrical}. In this paper, we focus on the interplay between the lunar regolith and foregrounds and leave an analysis of coupling between antennas to future study. Finally, foreground power is expected to turnover due to free-free absorption at low frequencies, though the exact frequency varies with foreground regions \citep{Seitova_2022}. This final factor is not expected to affect Epoch of Reionization (EoR) analysis at higher frequencies, but could have an impact on the Dark Ages window at or below 10 MHz. Beyond these factors is the inherent redshift dependence of the foreground wedge itself which drastically shrinks the available modes not contaminated by foregrounds at high redshifts and further necessitates a high-precision understanding of foreground contamination \citep{pober2025impactforegroundsdarkages}.

In this paper, we examine the level of foreground power in reference to the 21\,cm power spectrum in order to determine if a window of detection to the cosmic Dark Ages signal is possible. We follow a delay spectrum approach in our analysis along the lines of \cite{Parsons_2012b} and \cite{Lanman_2020} as described in Section~\ref{sec:theory}, and simulate the observation of various baselines under flat and spectrally varying sky models, along with both analytical and full-wave\footnote{One of the most accurate CEM methods that involves solving Maxwell's equations using propagation matrix methods and solutions involving plane wave expansions \cite[e.g.][]{nagano1975numerical, yin2019fullwave}.} computational electromagnetics (CEM) beam models using the commercial CST Studio Suite program described in Section~\ref{sec:simParam}. We analyze the output of these simulations in Section~\ref{sec:results}, where we take special note of the spectral structure imparted by both sky and beam models. Finally, we discuss the implications of these results in Section~\ref{sec:discuss} and how large foreground leakage power due to the complex beam spectral structure necessitates high accuracy models.

 Throughout this paper we assume a Planck 2018 cosmological model with $h=0.6766$ \citep{2020A&A...641A...6P, astropy:2013,astropy:2018,astropy:2022}. 

\section{Foreground Power} \label{sec:theory}

In this section, we describe the methodology we use to analyze the foreground power in roughly the same process as described in \cite{Lanman_2020}. We first simulate the 21\,cm power spectrum in the absence of foregrounds. This power spectrum can then be compared to the foreground power, which we estimate from simulated visibilities via a delay spectrum approach as described in \ref{subsec:delayTrans} and \ref{subsec:delayCalc}. Finally, we quantitatively describe the foreground power \textit{spillover}, which represents how much of the power spectrum measurement is dominated by foregrounds. 

\subsection{The 21 cm Power Spectrum} \label{subsec:powerSpec}

The typical object of study in 21\,cm cosmology is the 21\,cm power spectrum $P(\bf{k})$ which is defined via the two-point function of brightness temperature fluctuations as:

\begin{equation}
    \langle T_{b}({\bm{k}}) T_{b}({\bm{k'}}) \rangle
    = (2\pi)^3 \delta ^3({\bm{k}} - {\bm{k'}})P_{21}(\bm{k})
    \label{twopoint},
\end{equation}

\noindent where $\bf{k}$ is the wave vector. $T_{b}(\bf{k})$ is the brightness temperature, which is observed via the 21\,cm signal, and it traces the density of \HI along with the ionization fraction, spin temperature, and velocity gradient along the line of sight.

In our analysis, we generate two $\Lambda$CDM 21\,cm brightness temperature power spectra in \texttt{21cmFast} \citep{greig_and_mesinger_2018} using the default settings with spin fluctuations turned on: one corresponding to $z=30$ \citep[the same one used in][]{smith2025detecting21cmsignal} and one for $z=70$. These 3D power spectra are averaged along spherical shells of constant magnitude $|\textbf{k}|$ to form 1D power spectra (although due to cosmological scalings, at Dark Ages frequencies the observed $|\textbf{k}|\approx k_{\parallel}$). These reference power spectra set the nominal threshold to which foreground power must be suppressed in order to observe the 21\,cm signal. 

\subsection{The Delay Transform} \label{subsec:delayTrans}

A radio interferometer does not directly measure $T_{b}(\bf{k})$ as it is constrained by a discretized baseline distribution and directional electric field ($E$-field) sensitivity. Rather, the measurement of a radio interferometer can be best described as a complex visibility:

\begin{equation}
    V(\bm{b},\nu) = \int A(\hat{s},\nu)T_{b}(\hat{s},\nu)e^{-2\pi \textit{i}((\hat{s}-\hat{s}_{p})\cdot\bm{b})\nu/c}\diff^{2}s,
    \label{vis}
\end{equation}

\noindent where $\nu$ is the frequency, $\bf{b}$ is the baseline vector, $A$ is the primary beam describing the directional sensitivity of the antennas, and $\hat{s}_{p}$ is a unit vector pointing to the phase center. This integral is carried out over the entire sky. In a flat-sky approximation, baseline position $\bm{u}=\bm{b}\nu/c$ is the Fourier dual of direction cosines pointing from the phase center $\hat{s}$ which describes the position of a source on the sky.

There are typically two overarching methods that can be used to estimate the cosmological power spectrum: image reconstruction and the delay spectrum. Both methods ultimately involve a Fourier transform of the visibilities over some line-of-sight mode albeit with subtle differences \citep[see][]{10.1093/mnras/sty2844}. In this paper, we opt for the delay spectrum approach which involves a per-baseline Fourier transform over the frequency domain. While a full image-based power spectrum estimation is outside the scope of this paper, we discuss how the delay spectrum provides some intuition about other power spectrum estimators in Section \ref{sec:discuss}.

The delay spectrum takes the form:

\begin{equation}
    \tilde{V}(\bm{b},\tau) = \int\int \phi(\nu)A(\hat{s},\nu)T(\hat{s},\nu)e^{-2\pi \textit{i}(\tau_{g}-\tau)\nu}\diff^{2}s\,\diff\nu,
    \label{delay}
\end{equation}

\noindent where $\phi(\nu)$ is a spectral window function and $\tau$ represents the delay, the Fourier dual to frequency. $\tau_{g}=\hat{s}\cdot \bm{b}\nu/c$ is the \textit{geometric delay} which is the time it takes for light from a source at $\hat{s}$ to cross a baseline $\bm{b}$ --- i.e., $\tau_g=0$ for a source directly overhead and $\tau_g=|\bm{b}|/c$ for a source on the horizon parallel to the baseline. Whereas $\bf{u}$ is the Fourier dual to $\hat{s}$ and can therefore probe the perpendicular modes on the sky, $\bm{k}_\perp$,  $\tau$ is dual to $\nu$ which corresponds to redshift and therefore probes the line of sight Fourier modes $k_\parallel$ .

From Equation~\eqref{delay}, we can construct an estimation of the power spectrum with proper cosmological scaling \citep[][Appendix B]{Parsons_2012b, Parsons_2014}:
\begin{equation}
    P(\bm{k}_{\perp},k_{\parallel})\approx\frac{X^{2}Y}{B_{pp}\Omega_{pp}}|\tilde{V}({\tau})|^{2},
    \label{powerSpec}
\end{equation}

\noindent where the factors in the front approximate a cosmological volume: $X$ and $Y$ are redshift-dependent cosmological scaling factors evaluated at the center of the bandwidth and translate $\bm{u}$ and $\tau$ to  $\bm{k}_\perp$ and $k_\parallel$ respectively. $B_{pp} = \int |\phi(\nu)|^{2}\diff\nu$ is the effective bandwidth, and $\Omega_{pp}=\int|A(\hat{s})|^{2}\diff^{2}s$ is the square of the primary beam field of view where $A(\hat{s})$ is normalized to its peak value at each frequency.

\subsection{Delay Spectrum Calculation}\label{subsec:delayCalc}

In order to calculate the delay spectrum, we first simulate the visibilities, Equation~\eqref{vis}, in \texttt{pyuvsim} which we briefly describe in Section \ref{sec:simParam}.  From there, it is relatively straightforward to preform the discrete Fourier transform over frequency as shown in Equation~\eqref{delay} and to estimate the power spectrum in Equation~\eqref{powerSpec}.

However, the choice of spectral window $\phi(\nu)$ is not immediately obvious. A typical window function used in 21\,cm and radio astronomy is the Blackman-Harris. The Blackman-Harris window, described by a sum of cosines, is specifically designed to suppress side-lobes\footnote{Note, although they are unrelated, the term side-lobes is used both to refer to regions in the spectral leakage pattern and in the primary beam pattern, where here we refer to the spectral leakage pattern.} in Fourier space by $\sim$120~dB in power \citep{1455106}. The Blackman-Harris window is used in EoR analysis, where it is sometimes squared to provide an additional $\sim$120~dB of side-lobe suppression \citep{2016ApJ...825....9T}. However, we choose not to use this window function in our analysis, instead opting for a Dolph-Chebyshev window function which offers modularity in the dB power level of suppression --- where we use 180~dB and 240~dB \citep{1975tads.book.....R}. This choice of window function allows us to highlight the effects of spectral structure in the side-lobes which we discuss in Section~\ref{sec:results}.

\subsection{Foreground Spillover}
\label{subsec:spill}

In addition to analyzing spectral structure in the side-lobes, we seek to quantify the extent to which the foreground power main lobe spills over in delay space. If we consider spectrally flat power from a point source --- i.e., the window function, the beam, and the source brightness are independent of frequency --- with an infinite bandwidth, then we expect the power to reduce to a delta function centered at some geometric delay. The maximum geometric delay of this source occurs at the horizon, which is given in cosmological Fourier space as

\begin{equation}
    k_{\parallel,\text{horizon}} = \frac{X}{\nu Y}|\bf{k_{\perp}}|.
\end{equation}

\begin{table*}[htbp]
    \centering
    \begin{tabular}{lcc}
    \hline
        \textbf{Sky Models} & \\
        GSM Spectral Index & $n=-2.5$ with $\nu_\mathrm{ref}=45$\,MHz \\
        GSM Flat & Flat sky defined at $\nu_\mathrm{ref}=45$\,MHz\\
        \textbf{Beam Models} & \\
        Short Dipole & Modeled as a Hertzian dipole\\
        Achromatic Gaussian & $\sigma_{0}\sim0.5$\\
        Airy Disk & $D=6.18$\,m such that $A_\mathrm{eff}=30$\,m$^2$\\
        \textit{FarView} CST & Simulated with regolith models of 4 layers and 1 layer\\
        \textbf{Baselines} & \\
        11.5m Baseline & This is the spacing between two \textit{FarView} dipoles\\
        Line & Line of East-West baselines from 7\,m to 105\,m in 1\,m increments \\
        \textbf{Bandwidths} & \\
        10--28\,MHz & $140\gtrsim z \gtrsim 50$ \\
        32--50\,MHz & $44\gtrsim z \gtrsim 27$\\
    \hline
    \end{tabular}
    \caption{A summary of simulation parameters.}
    \label{SimParams}
\end{table*}

Thus, this horizon line defines the separation between spectrally smooth and spectrally structured power. While foreground power at Dark Ages frequencies is largely understood to be smooth \citep{10.1111/j.1365-2966.2003.07133.x} and the 21\,cm power is expected to be spectrally structured, the finiteness of bandwidth of a real observation means that foreground power will spread out and exhibit \textit{spillover} beyond the horizon line. Additionally, in order to suppress side-lobe levels below the Dark Ages 21\,cm power level, window functions like the Dolph-Chebyshev are necessary, but come with the unfortunate cost of expanding the main-lobe and increasing this foreground power spillover.

To quantify foreground power spillover, we define $k_{\text{spill}} = \text{max}(k_\parallel|_{P(k_\parallel)=P_{21}})$, i.e., the maximum $k_{\parallel}$ of the intersection between the delay spectrum estimate of the foregrounds and the $\Lambda$CDM 21\,cm power spectrum. This value thus represents the lowest accessible $k_\parallel$ of the Dark Ages window. We show $k_\text{spill}$ in reference to $k_{\parallel,\text{horizon}}$ in Figure \ref{fig:AnalCompare}.  

\section{Simulation Parameters} \label{sec:simParam}

In order to simulate foreground power, we use the comprehensive radio interferometer simulation package \texttt{pyuvsim} \citep{2019JOSS....4.1234L}. \texttt{pyuvsim} is specifically designed with the goals of 21\,cm cosmology in mind and offers robust and accurate end-to-end simulation results. \texttt{pyuvsim} accepts input parameters for frequencies and bandwidth, source catalogs of varying spectral types, full telescope configurations including layout and beam models, and observation times. It then uses a full Jones-matrix formalism to add contributions from sources to each baseline time taking into account the antenna's $E$-field primary beam. Finally, these contributions are integrated over the full solid angle of the sky to construct visibilities for each baseline, time and frequency as in Equation~\eqref{vis}. In this section, we describe the simulation parameters we use in \texttt{pyuvsim} which are summarized in Table~\ref{SimParams}. 

\subsection{Sky Models} \label{subsec:skyModel}

In EoR analysis, the choice of sky model is a nuanced decision. However, the low frequency range of the Dark Ages, where sky maps become fewer and sparser, constrains the choice of sky model. For our analysis, we use the 2016 Global Sky Model (GSM) which is constructed from 29 sky maps between 10\,MHz and 5\,THz \citep{2017MNRAS.464.3486Z}. Admittedly, this restricts our analysis in that the GSM cannot extrapolate below 10\,MHz. However, 10\,MHz corresponds to $z\sim141$ which, given that the Dark Ages 21\,cm signal ceases to be observable around $z\sim 200$ \citep{Furlanetto_2006}, gives a sufficient lower limit on frequency. We then impose an upper limit on frequency of 50\,MHz which corresponds to $z\sim27$.

One caveat for using the GSM at low frequencies is the sparseness and incompleteness of sky maps below 50\,MHz. This incompleteness contributes unexpected spectral structure to the GSM at low frequencies which is not present in the EoR ($\nu\gtrsim100$\,MHz) delay spectrum analysis in \cite{Lanman_2020}. From \cite{2017MNRAS.464.3486Z}, a larger rms error arises below 100\,MHz, peaking at 30\,MHz, likely from the sparseness of low frequency sky maps. Therefore, it is difficult to discern true spectral structure from spectral artifacts. However, galactic foregrounds are expected to be spectrally smooth above 10\,MHz\footnote{A small percentage of optically thick regions will exhibit a turnover in foreground power due to free-free absorption, but current $>1\sigma$ confidence estimates suggest $<1\%$ of the sky will exhibit this turnover \citep[see][]{Seitova_2022}.}, and it is this fact that we aim to exploit in Dark Ages observations. Thus, we look to avoid unwanted spectral structure in our sky model and opt to use the most substantially mapped out frequency in the GSM, 45\,MHz, and apply an average spectral index of -2.5 --- the power law index of the dominant galactic synchrotron radiation \citep[e.g.,][]{bowman2018absorption} --- to map the remaining frequencies. This method leaves the intrinsic frequency dependence of the sky in our model while minimizing any potential artifacts or odd frequency structure imparted by the GSM. Of course, if the spectral artifacts observed in the GSM at low frequencies are truly representative of the radio foregrounds, 21\,cm cosmology will be significantly more challenging; in this regard, our analysis can be considered optimistic.

Additionally, we can further simplify our sky model by removing the spectral index and broadly applying the 45\,MHz sky to every frequency. This second, flat sky model allows us specifically observe effects of frequency dependence in the beam only.

\subsection{Beam Models} \label{subsec:beamModels}

\subsubsection{Analytic Beams} \label{subsubsec:analyticBeams}

We begin with a number of analytical beam models that give a basis of behavior for the visibilities captured by our instrument. Analytical beam models do not take into account the lunar regolith or the true size and shape of the antenna. We divide these models into two categories: chromatic (with frequency dependence) and achromatic, or flat (without frequency dependence). Each of these analytic beam models is conveniently wrapped into \texttt{pyuvdata} which interfaces with \texttt{pyuvsim} \citep{Hazelton2017, 2025JOSS...10.7482K}.

We consider one chromatic analytic beam model since the \textit{FarView} beam model is itself chromatic: the \textit{Airy} beam. The \textit{Airy} beam is defined as the far-field radiation pattern of a uniform circular antenna, which takes on the form of an Airy disk:

\begin{equation}
    A_\mathrm{Airy} = \frac{\nu/c}{\pi D\sin\theta}J_{1}\left(\frac{\pi D\sin\theta}{\nu/c}\right),
\end{equation}

\noindent where $\theta$ is the zenith angle, $J_{1}$ is the first-order Bessel function of the first kind, $D$ is the diameter of the antenna, $\nu$ is frequency, and $c$ is the speed of light. Here we choose $D=6.18$\,m such that $A_\mathrm{eff}$ of the dish is equal to $A_\mathrm{eff}=30$\,m$^2$ of a \textit{FarView} dipole.

While the Airy beam approximates the collecting area of a \textit{FarView} dipole, it does not take into account its dipole shape. Let us first consider the quintessential \textit{short dipole}. A \textit{short dipole} is defined as having a total length significantly less than half the wavelength of the observational frequency. For our analysis, which is confined between 10 and 50 MHz, the largest wavelength is 30\,m. The \textit{FarView} dipole is 10\,m in total length, meaning that a short dipole beam does not completely approximate the \textit{FarView} beam. However, a short dipole is useful in that it provides a somewhat analogous picture albeit without chromaticity. A short dipole beam takes the form of:

\begin{equation}
    A_\mathrm{SD}(\theta) = \sin^{2}\theta.
\end{equation}

We also consider an \textit{achromatic Gaussian} beam, which is defined as a Gaussian with a constant full width at half-maximum (FWHM) at all frequencies:

\begin{equation}
    A_\mathrm{Gauss}(\theta) = \exp\left(-\frac{\theta^2}{2\sigma_{0}^{2}}\right), 
\end{equation}

\noindent where $\sigma_{0}$ can be tweaked to set the FWHM of the achromatic Gaussian. While an achromatic Gaussian is more applicable to a circular dish antenna, the modularity of $\sigma_{0}$ allows us to encode some information about the size of the antenna. In this case, we set $\sigma_{0}$ to the width of the main lobe of the Airy beam.

\subsubsection{Numerically simulated \textit{FarView} antenna} \label{subsubsec:farview}

For comparison with a more realistic crossed dipole antenna beam as would be implemented by \textit{FarView} --- or other lunar far side arrays planning to use crossed dipole antennas such as CoDEX/DEX \citep{koopmans2019peeringdarkageslowfrequency, brinkerink2025darkagesexplorerdex} --- we utilize the commercial CEM software, the CST Studio Suite\footnote{\url{https://www.3ds.com/products/simulia/cst-studio-suite}}, to simulate the farfield radiation beam patterns of a \textit{FarView} antenna between 5--50~MHz. The \textit{FarView} antenna element's full-wave beam models were computed by the CST's transient time-domain solver using the Finite Integration Technique (FIT)\footnote{Due to the much faster simulation time achieved by the FIT solver, a variant of the Finite Difference Time-Domain (FDTD) solver, over the entire frequency range in time domain with GPU acceleration, we have only spot checked our FIT results with CST's slower Method of Moments (MoM) frequency-domain solver, which solves one frequency per realization. Since the beam models here are not meant to be compared to an actual instrument, maintaining self-consistency between the 1 and 4 layer models with identical FIT solver settings and boundary conditions is more important.} \citep{davidson210cem}. The CST simulation was setup to run between 0--60 MHz, and output a 3D far-field beam in spherical coordinates at every 125~kHz\footnote{This resolution is selected to be fine enough to see high $k_{\parallel}$ structure in delay space.} between 5--50~MHz.

\begin{figure}[ht!]
    \centering
    \includegraphics[width=1.0\columnwidth]{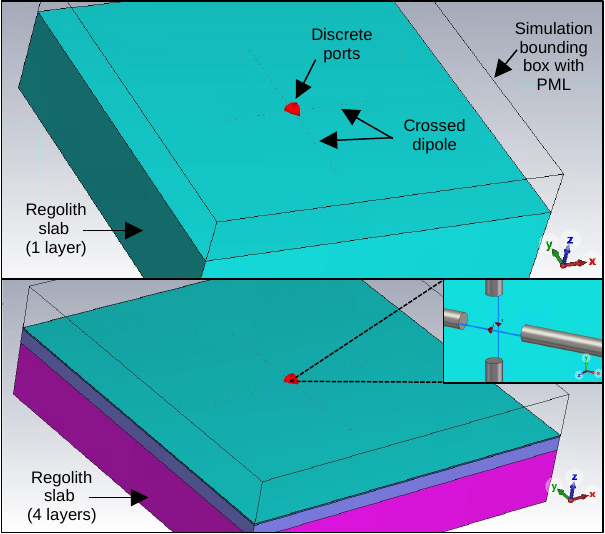}
    \caption{The 3D CAD model constructed in CST for the two layered-regolith versions: 1 layer \textit{(top)}, 4 layers \textit{(bottom)}. The inset shows the identical discrete port configuration for the dipole pair for both models, with Port 1 (or 2) designated for the dipole along the $x$ (or $y$) axis in the local coordinates.}
    \label{fig:farview_cst_cad}
\end{figure}

\begin{table}
    \begin{tabular}{|l|c|c|c|} 
        \hline
        Layer No. & $\epsilon_r$ & $\tan(\delta)$ & Thickness [m] \\ 
        \hline
        1 (Top) & 3.00 & 0.02   & 0.01 \\
        2 & 3.25 & 0.01   & 0.09 \\
        3 & 3.50 & 0.0125 & 0.91 \\
        4 (Bottom) & 3.75 & 0.15   & 4.09 \\
        \hline
    \end{tabular}
    \caption{CST model material properties for the 4-layer regolith slab.}
    \label{tbl:cst_para_4layers}
\end{table}

\begin{figure}[htb!]
    \centering
    \includegraphics[width=1.0\columnwidth]{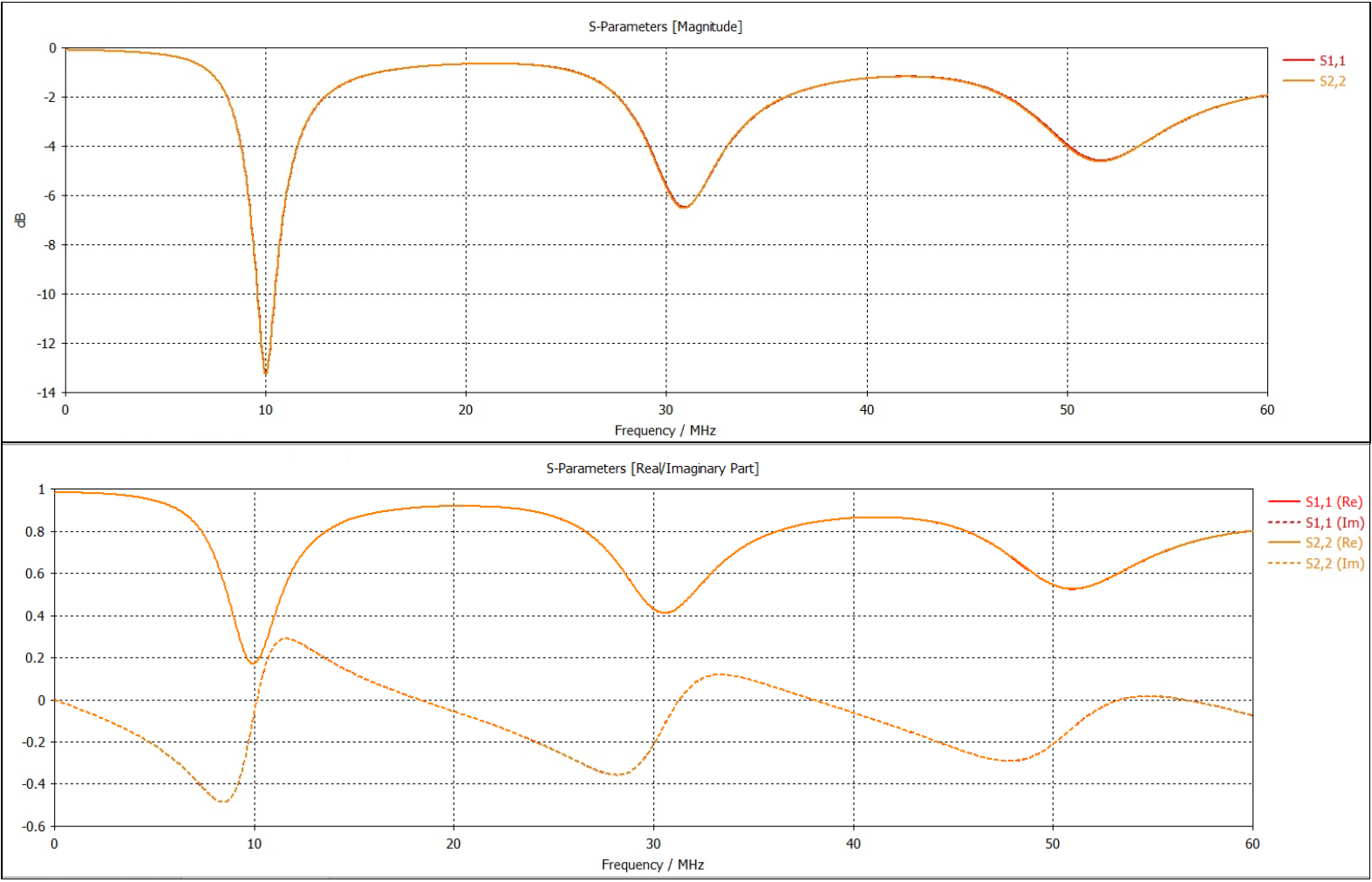}
    \caption{Antenna reflection coefficients, in terms of $S_{11}(\nu)$ (Port 1) and $S_{22}(\nu)$ (Port 2), for the 4-layer \textit{FarView} model shows the impedance match (relative to a 50~$\Omega$) and resonant frequencies of the cross dipole: Magnitude value in dB \textit{(top)}, Real and Imaginary parts \textit{(bottom)}. Note that the reflection coefficient of the two ports should near identical. The only slight apparent difference is due to the placement offset of the discrete ports in CST for the two dipoles in the simulator to ensure the two ports are not overlapping (see inset in Figure~\ref{fig:farview_cst_cad}). This has negligible effects in this study.}
    \label{fig:cst_farview_s11_04layer}
\end{figure}

The \textit{FarView} element was modeled as a pair of simple crossed-dipole antenna with arm length of 5~m with a small dipole gap of 20~cm (i.e., tip-tip length of 10.2 m), and a dipole diameter of 5~mm \citep[based on][]{polidan2024farviewinsitumanufacturedlunar}. The dipole pair is situated on the top surface of the regolith substrate in the 3D CAD (Figure~\ref{fig:farview_cst_cad}). The regolith underneath is modeled as a slab of dielectric material with sides of 20~m\footnote{This value of the soil slab side for the antenna element is chosen for the convenience of future array simulation with duplicated elements placed next to each other with an antenna separation of 11.5~m, one of the preliminary design parameters.} at a certain thickness and using the Perfectly Matched Layer (PML) boundary condition method to approximate an infinite regolith at the interfaces with the CST FIT's bounding box (in $\pm x$, $\pm y$, and $-z$ directions, where the origin of the coordinates is the center of the dipole gap). Additionally, the simulation for each of the dipole polarization is run sequentially by exciting the two CST $S$-parameter type discrete ports\footnote{This type of CST port is modeled as two thin wires with a current source (red cone symbol in Figure~\ref{fig:farview_cst_cad}) with inner impedance that excites and absorbed a signal power of 1~Watt, commonly used for time-domain transient antenna simulation.} placed in the dipole gaps with antenna port impedance set at 50~$\Omega$.

\begin{figure*}
    \gridline{
    \includegraphics[width=0.65\linewidth,valign=c]{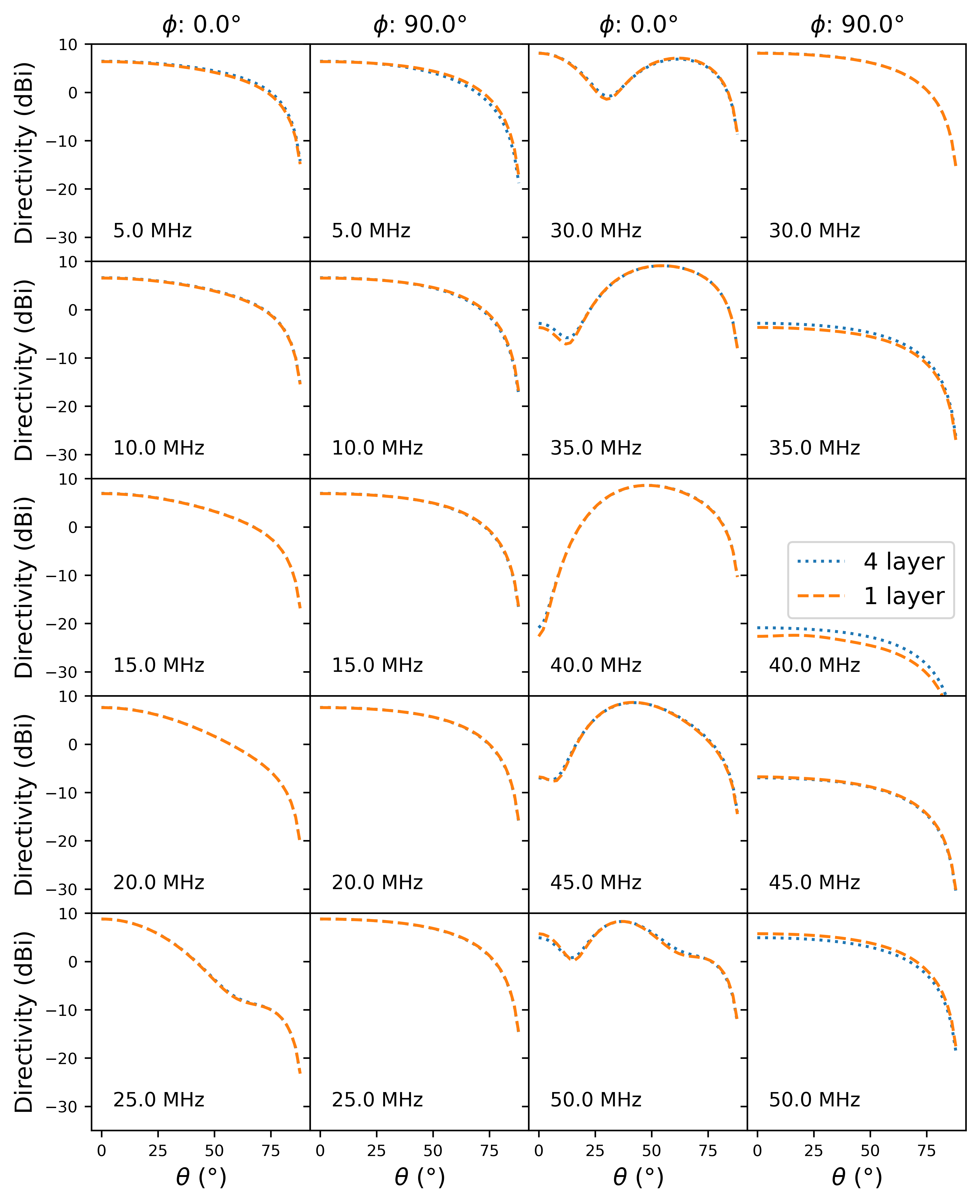}
    \includegraphics[width=.33\linewidth,valign=c]{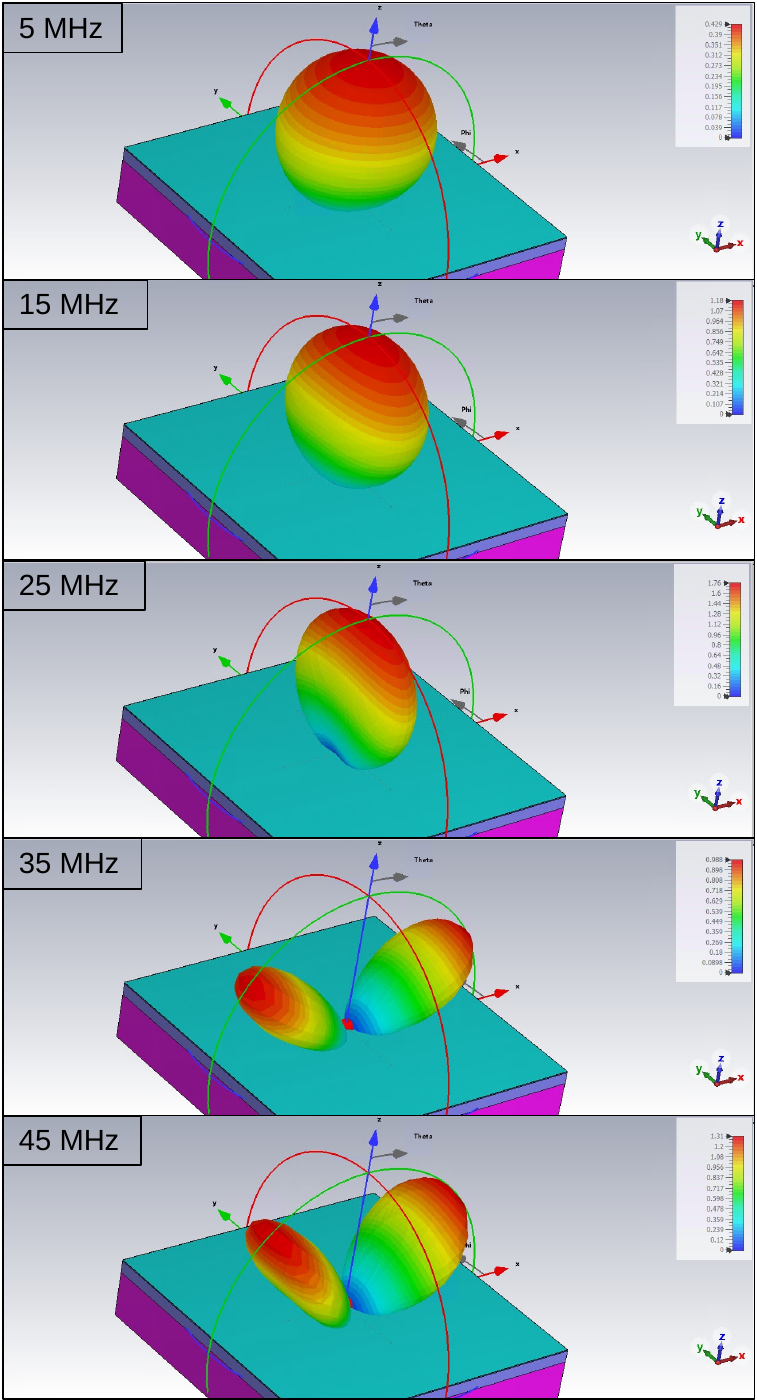}
    }
    \caption{\textit{(Left)} Comparison of a single \textit{FarView} dipole beams at the two principle planes ($E$- and $H$-planes at $\phi=0^{\circ}$ and $90^{\circ}$, respectively) in terms of beam directivity in dB scale between the 1-layer (orange) and 4-layer models (blue) at every 5~MHz between 5--50~MHz. \textit{(Right)} Representative 3D beam directivity patterns (in linear scale) of the 4-layer \textit{FarView} CST model plotted at every 10~MHz showing the topological change in the beam as a function of frequency as the dipole response varies about its resonant frequencies. Since the dipole antenna intrinsically is a narrow-band resonant antenna, the far-field beams will go through phase changes as it moves away from the optimal resonant frequency (around 10~MHz for our 5-m dipole). This is evident by how the beams start bifurcating above 30~MHz, where the main lobe is no longer at the boresight ($\theta = 0^{\circ}$).}
    \label{fig:cst_farview_beam_plots}
\end{figure*}

The lunar regolith was modeled in two ways to illustrate underlying effects of a depth-dependent regolith. Firstly, a single layer of 5.0~m thick slab with an effective dielectric constant (or relative permittivity) $\epsilon_r = 3.0$ and a loss tangent $\tan(\delta) = 0.02$ is used as the fiducial beam. Secondly, based on previous studies on lunar regolith EM properties as a function of depth \citep[e.g.,][]{olhoeft1975dielectric,heiken1991lunar,li2022lunar}, a simplified four-layer regolith model was constructed with the parameters shown in Table~\ref{tbl:cst_para_4layers}, with the bottom layer treated with infinite depth with the PML boundary conditions. 

\begin{figure*}[htb!]
    \centering
    \includegraphics[width=0.85\linewidth]{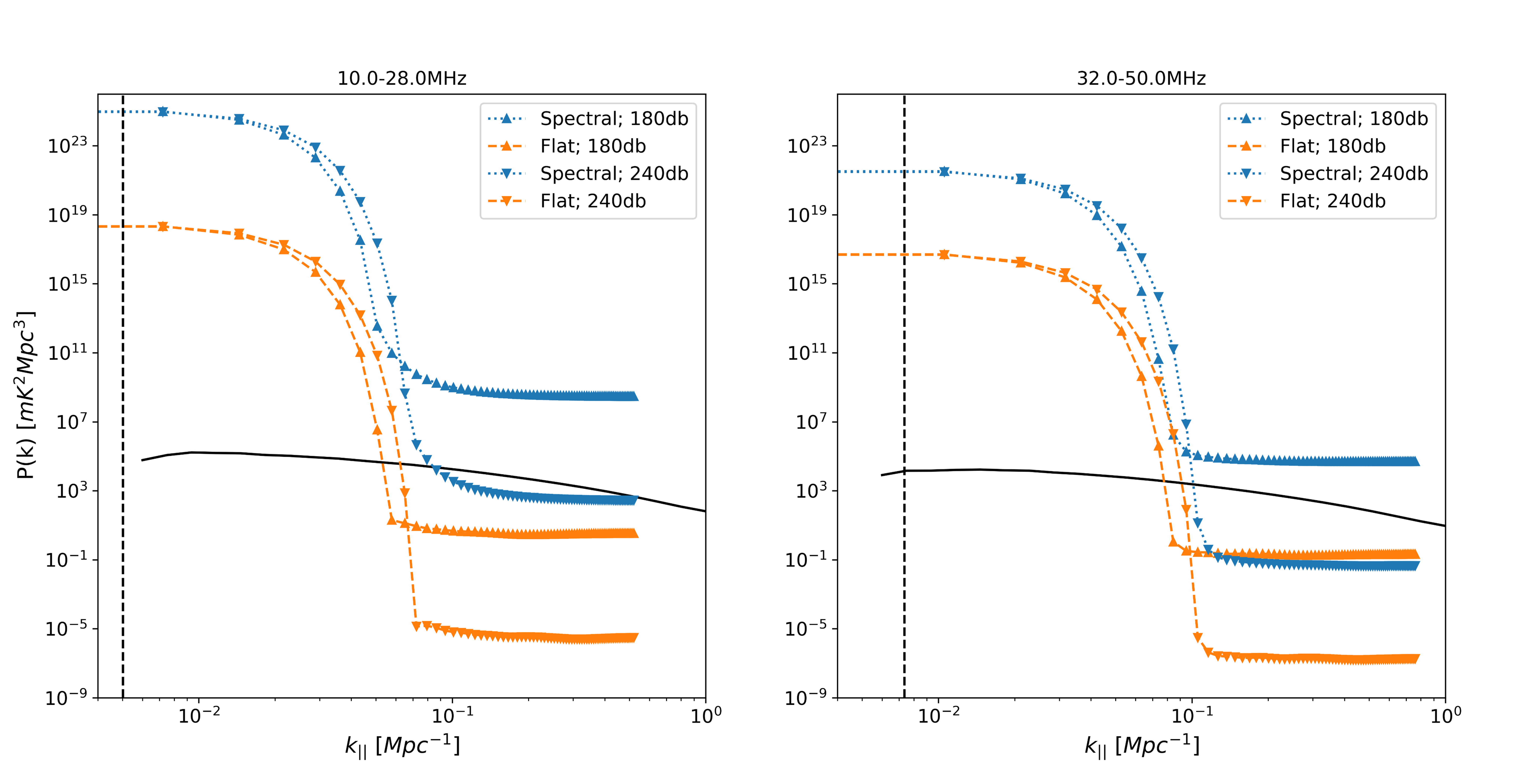}
    \caption{A comparison of various foreground power delay spectra for the Airy beam analytic model at bandwidths 10--28\,MHz (left) and 32--50\,MHz (right). The blue dotted lines show the delay spectra calculated from visibilities using the spectral index GSM and the orange dashed lines show the flat sky GSM. Upright and inverted triangles represent 180~dB and 240~dB Dolph-Chebyshev window functions respectively. The delay spectra are compared to the $z=70$ (left) and $z=30$ (right) 21\,cm power spectrum (solid black line); modes where the foreground power falls below this line are, in principle, detectable without any foreground subtraction. Finally, the vertical black dashed line shows the $k_{\parallel}$ of the horizon. The 21\,cm power spectrum and horizon lines appear in subsequent delay spectra figures under the same convention.}
    \label{fig:AiryCompare}
\end{figure*}

The three resonances frequencies (around 10, 21, and 52~MHz) are apparent in the antenna reflection coefficient, as the scattering parameter $S_{11}(\nu)$ (Port 1) and $S_{22}(\nu)$ (Port 2), for the 4-layer model is shown in Figure~\ref{fig:cst_farview_s11_04layer}. Since the dipole is intrinsically a narrow-band resonant antenna, the far-field beams will go through phase changes as it moves away from the optimal resonant frequency and its harmonics. As a result, the far-field beam pattern will change, including bifurcating into two main lobes at an offset angle instead of a single main lobe at the boresight direction. Without the loss of generality, Figure~\ref{fig:cst_farview_beam_plots} shows a comparison of subsequent far-field beams for the $x$-orientated dipole between the 1-layer and 4-layer models at every 5~MHz, along with a subset of representative 3D beam patterns for the 1-layer model to illustrate the beam variations, including how the beam bifurcation behavior.

\subsection{Additional Simulation Parameters}\label{subsec:addParams}

To complete our realistic simulation of foreground power as observed on the lunar surface, we must define an array configuration and finite bandwidth. For the array configuration, we focus on two baseline layouts. We first simulate a pair of antennas forming a single 11.5\,m baseline which is representative of the spacing between two \textit{FarView} dipoles. In practice, \textit{FarView} is intended to use subarrays of 16 dipoles beamformed together, with a typical separation between subarrays of 49\,m. However, we opt not to simulate these subarrays. Instead, in a more general application to cover a sufficient breadth of baseline lengths, we additionally simulate a line of East-West antennas to form a range of baselines from 7\,m to 105\,m in 1\,m increments. Longer baselines, which correspond to larger $k_\perp$, probe smaller angular scales on the sky and of interest to our simulations, offer a picture of how foreground spillover interacts with the wedge. Since mutual coupling between antenna elements is outside the scope, these simulations are simply duplicating the antenna beam across the array. Future CST simulations of a small \textit{FarView} array can help to constrain these mutual coupling effects better as illustrated by previous studies \citep[e.g.,][]{fagnoni2021design, josaitis2022array, rath2025investigating}.

For bandwidth, we choose an 18~MHz bandwidth, also representative of \textit{FarView} and of any Dark Ages experiment where large bandwidths (relative to the $\Delta z$ of the observation) are necessary to observe small $k$ where thermal noise is less dominant \citep{smith2025detecting21cmsignal}. 
We choose two bands for analysis: 10--28\,MHz and 32--50\,MHz.  As shown in Figure~\ref{fig:cst_farview_beam_plots}, the \textit{Farview} beam transitions from a single lobe with maximum directivity towards zenith to a bifurcated beam around $\sim30$\,MHz.  Avoiding this transition frequency in our  delay spectrum analysis simplifies the interpretation of the results for this specific analysis, although there is no fundamental reason these frequencies cannot be used in a cosmological analysis.

\section{Results}\label{sec:results}

\subsection{Analytic Beam Results}
\label{subsec:analResults}
In our analytic beam analysis, we examine the effects of chromaticity in the sky and beam models as described in Sections~\ref{subsec:skyModel} and \ref{subsec:beamModels}. Namely, we ``toggle'' chromaticity to produce four cases of study: (i) chromatic sky and chromatic beam, (ii) flat sky and chromatic beam, (iii) chromatic sky and flat beam, and (iv) flat sky and flat beam. For each case, we perform a delay spectrum analysis on visibilities simulated in \texttt{pyuvsim} for an individual 11.5\,m baseline, along with a line of East-West baselines as described in Section \ref{subsec:addParams}.

We first examine how a chromatic beam (the Airy beam) behaves against flat and spectral skies. In Figure~\ref{fig:AiryCompare}, delay spectra of the foreground power for spectral index and flat GSM skies are compared to Dark Ages 21\,cm power spectra at redshifts $z=70$ and $z=30$ for the 10--28\,MHz and 32--50\,MHz bandwidths respectively. A Dolph-Chebyshev window function is applied at 180~dB and 240~dB where 240~dB is necessary to bring the spectral index sky delay spectrum below the Dark Ages power level. Notably, despite the chromaticity of the Airy beam model, the Dark Ages window appears quite smooth for both sky models.

\begin{figure}[htb!]
    \centering
    \includegraphics[width=1.0\linewidth]{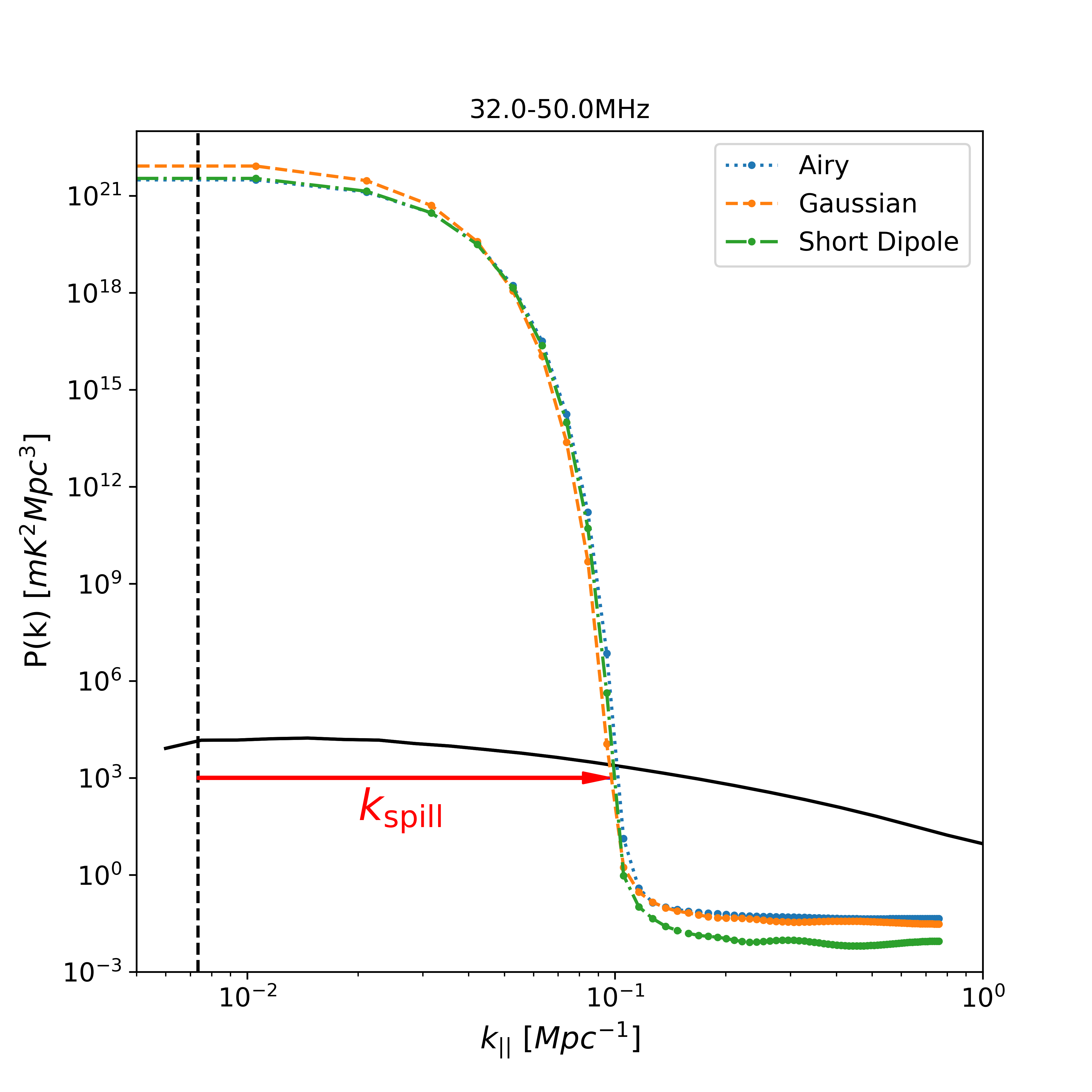}
    \caption{A direct comparison of delay spectra of analytic beam models (Airy: blue dotted; Gaussian: orange dashed; Short Dipole: green dot-dashed) against the spectral index GSM using a 240\,dB Dolph-Chebyshev, 11.5~m baseline, and 32--50\,MHz bandwidth. The red arrow points towards $k_{\text{spill}}$ for the Gaussian beam delay spectrum and shows the extent of spillover from $k_{\parallel,\text{horizon}}$}
    \label{fig:AnalCompare}
\end{figure}

Next, we compare the Airy beam to achromatic beams (Gaussian and Short Dipole beams) against a spectral sky. Overall, we find little difference between the chromatic and achromatic analytic beams for the 11.5\,m baseline. We show the delay spectra of the three simulated beam types with spectral index sky in Figure \ref{fig:AnalCompare}. In general, we find that, due to a lack of complex spectral structure, these analytic beam models present an optimistic window to the Dark Ages 21\,cm power spectrum. Additionally, the strength of foreground power at Dark Ages frequencies requires substantial window functions that come at the cost of larger foreground spillover values than EoR windows.

\begin{figure}[htb!]
    \centering
    \includegraphics[width=1.0\linewidth]{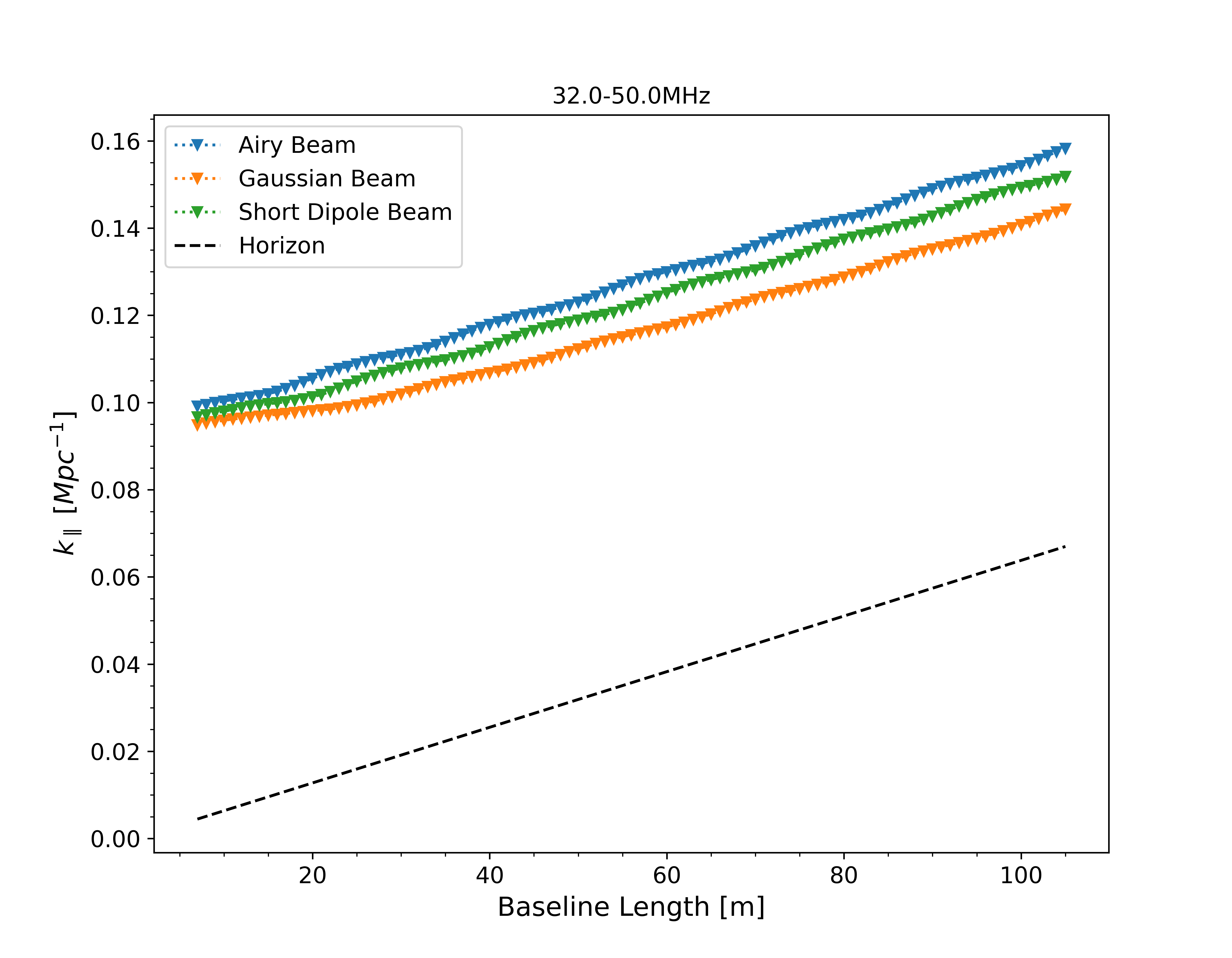}
    \caption{A comparison of foreground spillover values vs baseline length for the three analytic beam models (Airy: blue; Gaussian: orange; Short Dipole: green). The foreground spillovers were calculated from delay spectra using a 240\,dB Dolph-Chebyshev window and spectral index GSM. The horizon as a function of baseline length is shown as the black dashed line.}
    \label{fig:deltaCompare}
\end{figure}

We also extend the analysis of foreground spillover to more baselines using a line of 100 antennas with East-West baselines calculated between an antenna at the origin and each subsequent antenna between 7\,m and 105\,m. In Figure \ref{fig:deltaCompare}, we show the trend of $\text{max}(k_\parallel|_{P(k_\parallel)=P_{21}})$ with baseline length for each analytic beam. For longer baselines, we see an increase in the spillover and a visually distinct differentiation between analytic beam models. Oscillations present in Figure \ref{fig:deltaCompare} are a result of the evolving fringe pattern with baseline length.

\begin{figure*}[htb!]
    \centering
    \plotone{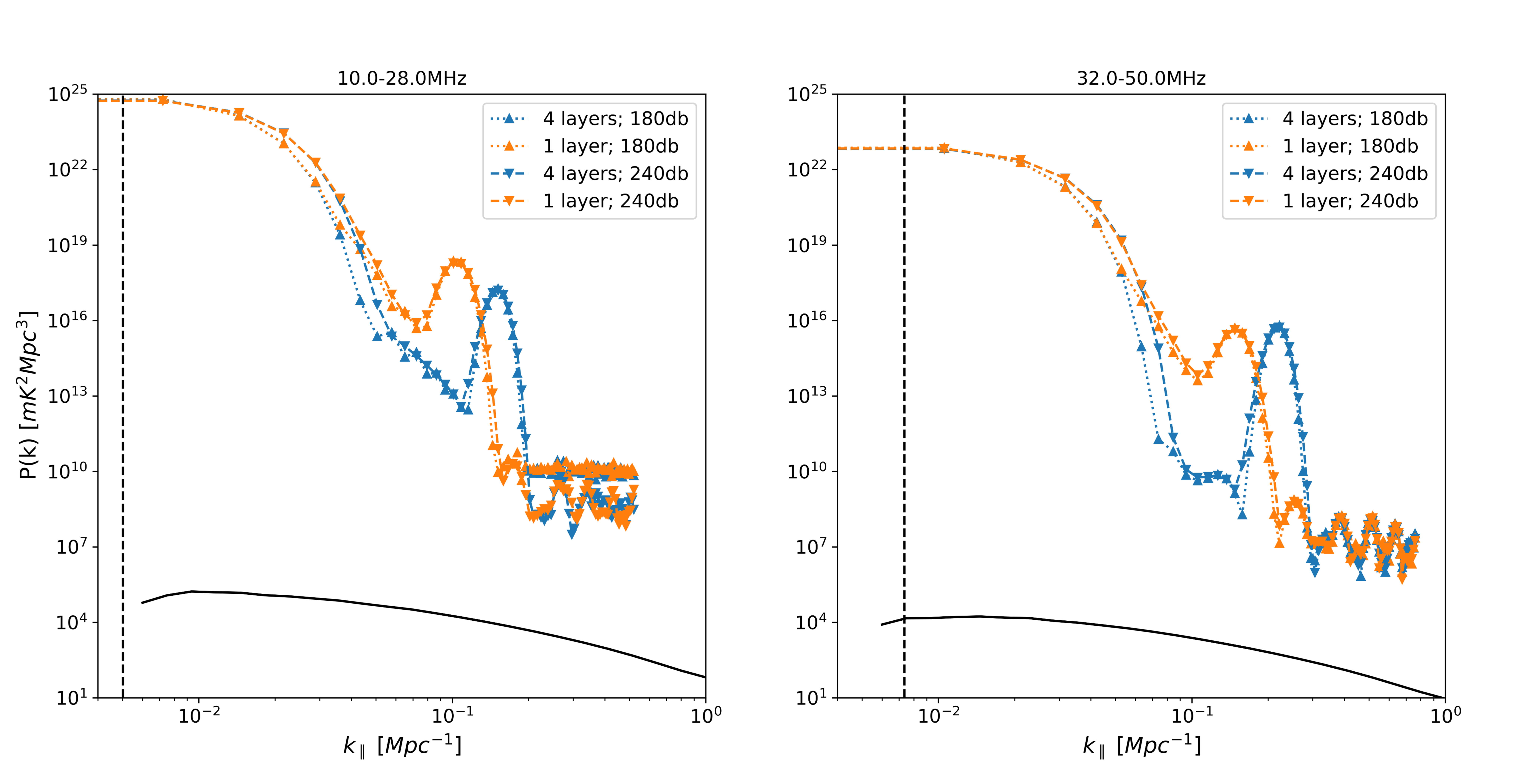}
    \caption{Delay spectra using the 4 layer (blue dotted) and 1 layer (orange dashed) lunar regolith \textit{FarView} beam models against a spectral index GSM for 10--28\,MHz (left) and 32--50\,MHz (right). Both 180\,dB and 240\,dB Dolph-Chebyshev window functions are used. Notably, higher decibel window functions do not bring the delay spectra side lobes below the 21\,cm power spectrum (solid black line) at $z=30$ (right) or $z=70$ (left), unlike the case for the analytic beams shown in Figure~\ref{fig:AiryCompare}.  This is indicative of true spectral structure in the beam, as opposed to an artifact of the signal processing.}
    \label{fig:CSTCompare}
\end{figure*}

\subsection{\textit{FarView} Beam Results}
\label{subsec:farviewResults}
We follow a similar approach in our analysis of the \textit{FarView} beam models as we do in the previous section. However, while the Airy beam model showed little difference to its achromatic counterparts, the chromaticity of the \textit{FarView} beam leads to complex spectral structure in the side-lobes of the foreground power. This structure varies with frequency and lunar regolith model as shown in Figure \ref{fig:CSTCompare}. Here, we plot delay spectra for the 1 layer and 4 layer lunar regolith \textit{FarView} beam models using a 180\,dB Dolph-Chebyshev window function. Notably, the delay spectra side lobes contain a large anomalous peak that decays into oscillations of similar height. This spectral structure lies several orders of magnitude above the Dark Ages power spectrum and sets the floor of the window --- using a higher decibel (dB) window function does not bring the delay spectra any lower.

\begin{figure*}[htb!]
    \centering
    \gridline{\includegraphics[width=1\linewidth]{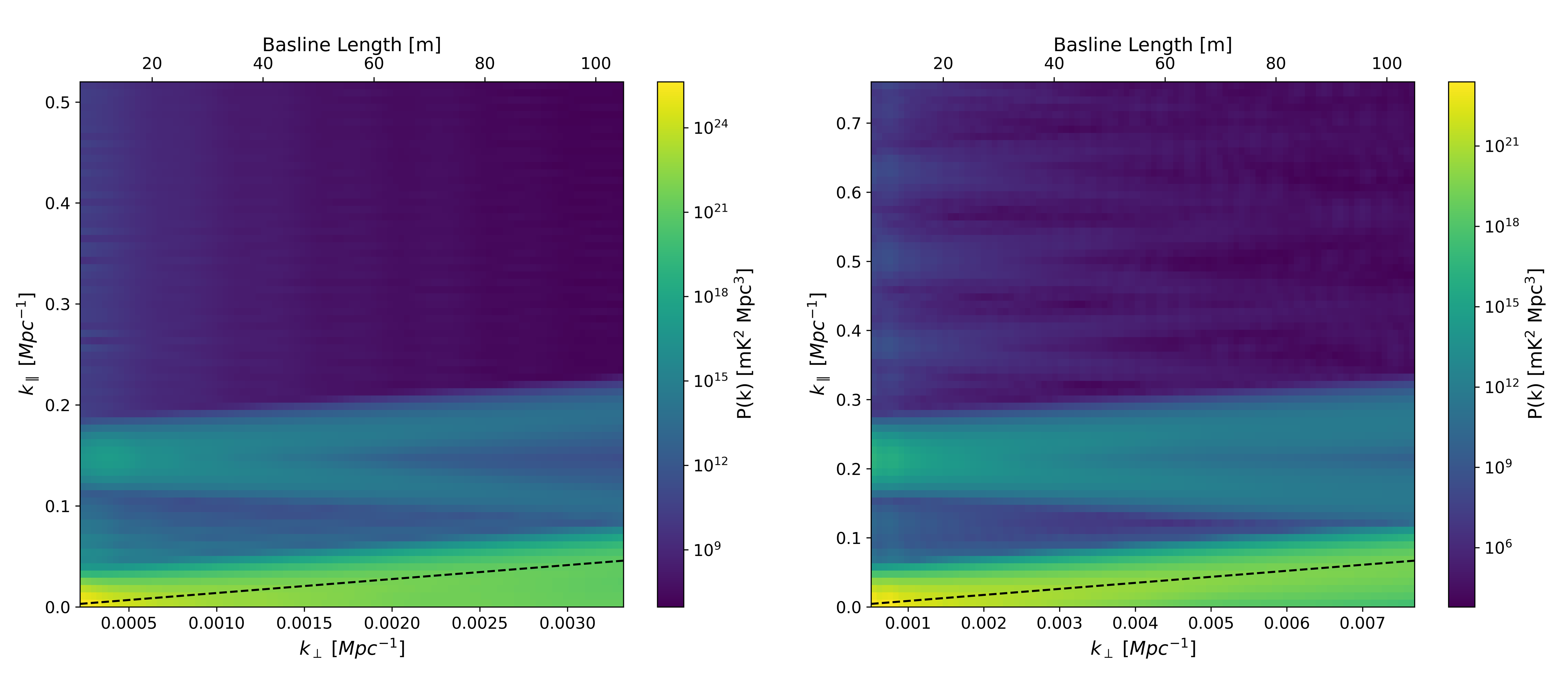}}{(a)}
    \gridline{\includegraphics[width=1\linewidth]{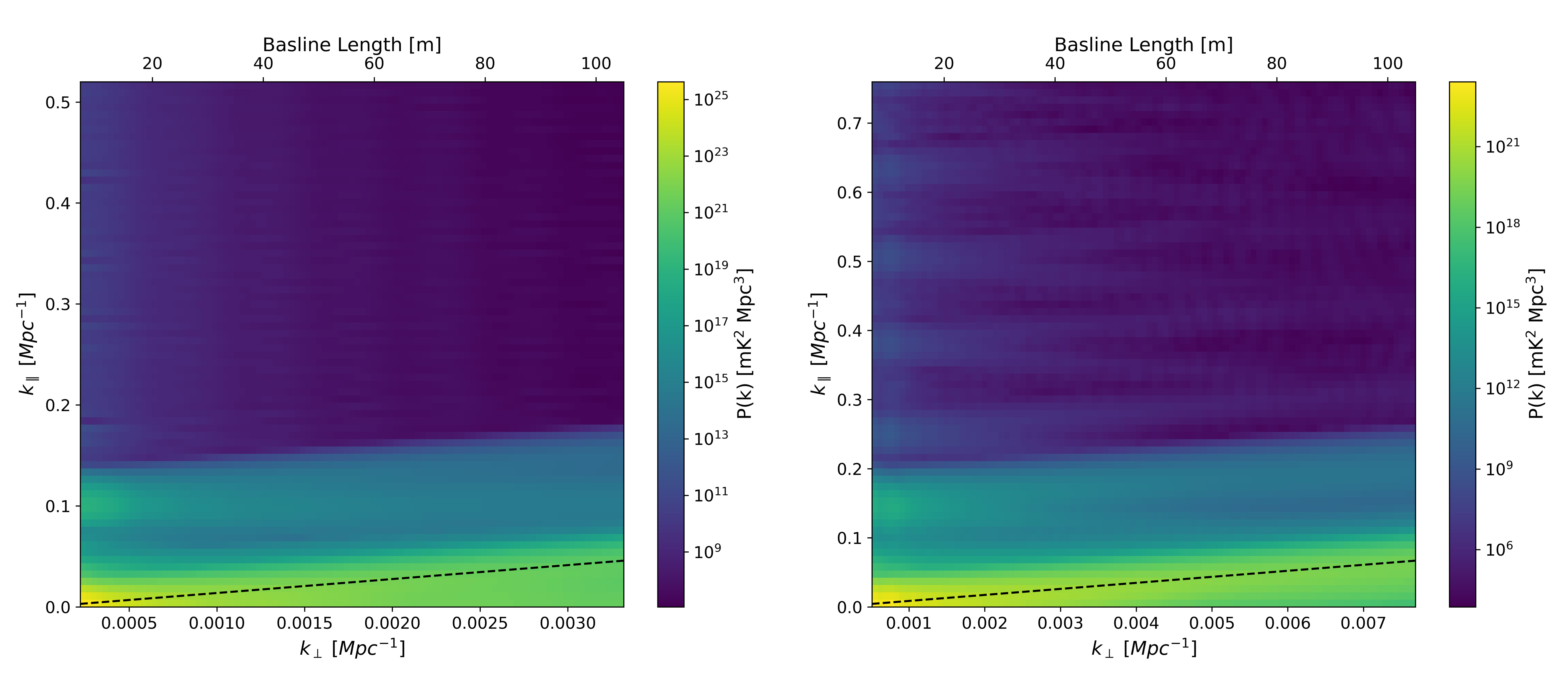}}{(b)}
    \caption{2D power spectra for \textit{FarView} beam models with (a) 4 layer regolith and (b) 1 layer regolith, against the spectral index GSM for a line of East-West baselines from 7\,m to 105\,m. The left and right columns are for 10--28\,MHz and 32--50\,MHz bandwidths respectively (Note: the color bar is at different scales per plot to emphasize low power features). At larger baselines it becomes clear that the foreground wedge is being duplicated, likely through reflection in the lunar regolith, at regularly spaced $k_{\parallel}$ above the $k_{\parallel, \mathrm{horizon}}$ (black dashed line).}
    \label{fig:FarView2d}
\end{figure*}

To better understand this structure, we simulate a line of East-West baselines using \textit{FarView} beam models to create 2D power spectra which we show in Figure \ref{fig:FarView2d}. In Fourier space, the 2D power spectra is mirrored in the negative $k_{\parallel}$ direction. Thus, it becomes clear at longer baselines, that the anomalous peak beyond $k_{\parallel, \mathrm{horizon}}$ is a duplication of the foreground wedge and its mirror. Subsequent oscillations are significantly lower power but a duplication of the same wedge. Recalling that $k_{\parallel}$ is proportional to delay, this effect suggests that foreground power is reflected at reduced power at regularly spaced delays. 

\begin{figure} [t!]
    \centering
    \includegraphics[width=1\linewidth]{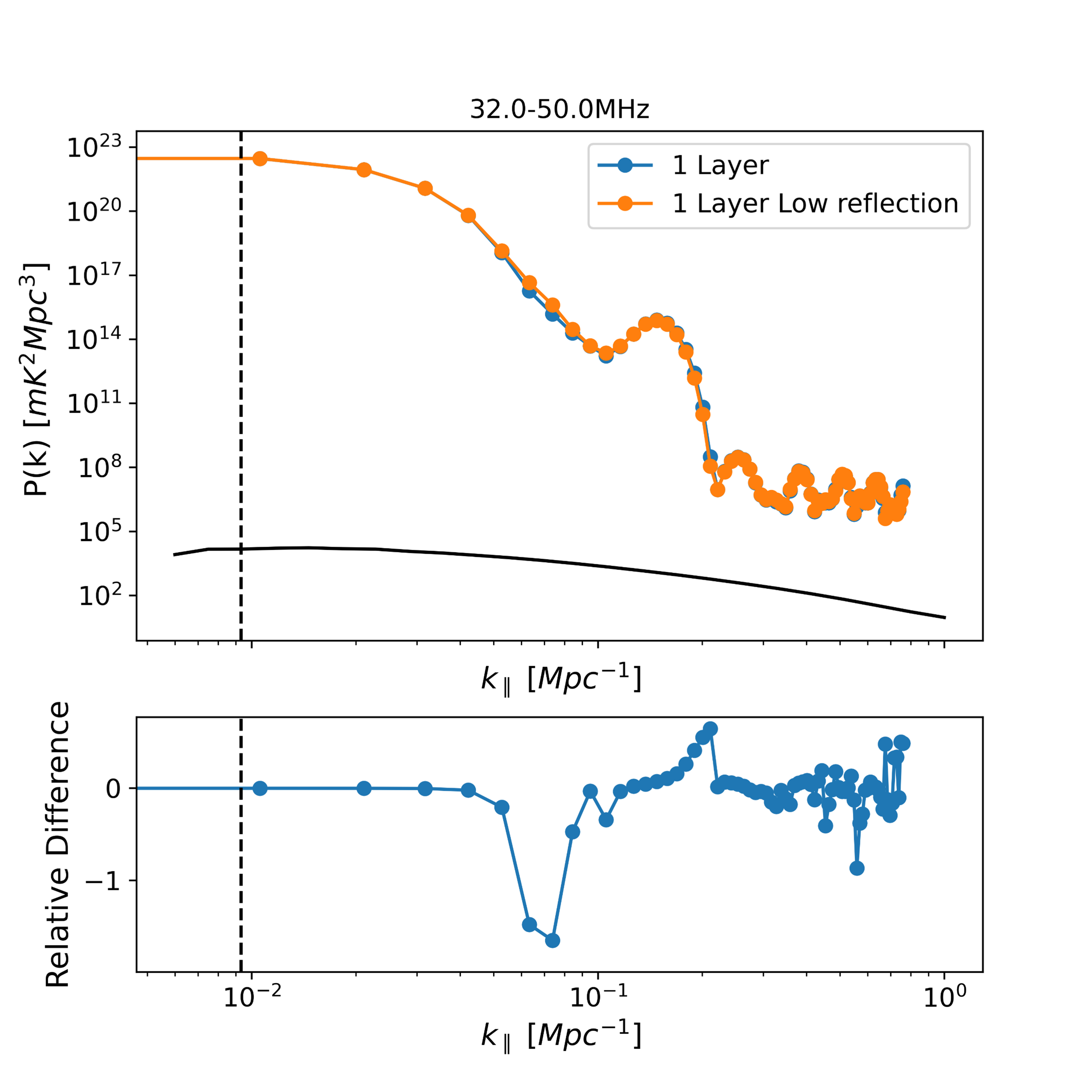}
    \caption{Delay spectra comparison for the 1-layer \textit{FarView} beam models at two different PML reflection levels: $10^{-4}$ (blue) and $10^{-6}$ (orange), along with their relative difference in the bottom panel. This suggests that it is unlikely the PML boundary condition has caused the unwanted spectral structure.}
    \label{fig:FarView2d_pml_compare}
\end{figure}

Since the oscillations are present in the delay spectra for both 1 and 4-layer models, even with position of the spectral structures shifted in $k_{\parallel}$, this indicates that they originate from the CST beams. One of the potential sources of oscillation is the PML boundary condition settings to emulate an infinite regolith layer. One of the PML parameters in CST is the estimated reflection level, which represents the potential residual radiation energy that gets reflected at the PML interface. The 1 and 4-layer models described above use a default PML reflection level of $10^{-4}$. As a sanity check, we also run the delay spectrum analysis with a third CST beam model using 1 layer of regolith with PML reflection level set to $10^{-6}$. However, no apparent changes are observed as shown in Figure~\ref{fig:FarView2d_pml_compare}. Therefore, we speculate that it is likely due to residual radiation in the simulation oscillating within the 5~m regolith slab in the current CAD model. However, an exploration of parameter space via a full suite of CST simulations to deduce the exact cause of this spectral structure is computationally expensive and exceeds the resources available.  We intend to explore this in future work, but for the purposes of the present analysis, the key conclusion is that the presence of the lunar regolith below the dipole introduces spectral structure in the beam and that the exact properties of that structure depend on the regolith model used.

\section{Discussion \& Conclusion}\label{sec:discuss}

Overall, we find that bright diffuse foreground emission makes achieving a window to the cosmic Dark Ages a difficult task. In the case of the analytic beam models, in which spectral structure is smooth, a window can be obtained. However, the combination of the bright foreground power, and resulting high-dB window function needed to deal with that power, along with a finiteness of bandwidth leads to substantial foreground spillover. From \cite{smith2025detecting21cmsignal}, thermal noise dominates the signal beyond $k\sim0.1\,\text{Mpc}^{-1}$ which corresponds roughly to $k_{\text{spill}}$ of the smallest simulated baseline (7\,m) using the Airy beam model at $z=30$. Furthermore, since foreground spillover only gets worse with longer baselines for all analytic beam models and remains consistently $0.1\,\mathrm{Mpc}^{-1}$ beyond the horizon, there is no conceivable window offered to the Dark Ages without foreground removal.

From the perspective of the more complex but realistic \textit{FarView} beam model, additional spectral structure introduced by electromagnetic interactions with the lunar regolith --- or other sources of spectral structure --- makes obtaining a window to the Dark Ages a challenging task. Unless the origin of the unwanted spectral structure and oscillations from the CST beam is understood, the foreground wedge dominates the signal and subsequent reflections set the floor of the Dark Ages window well above the 21\,cm power spectrum. 

While it may be possible to avoid coupling with the lunar regolith, as suggested in \cite{rotermund2026electrical}, by elevating the dipole antenna, doing this on the scale of an array the size of \textit{FarView} presents an incredibly difficult challenge. Alternatively, a design similar to DEX \citep{brinkerink2025darkagesexplorerdex} that uses inflatable antenna rigs may achieve a sufficient spacing to decouple from the ground, but further study of complex lunar regolith models with Dark Ages radio arrays is necessary to fully understand this phenomenon. Another option to avoid coupling, which has been used for arrays on Earth, is to place a ground screen below the entire array. This would effectively eliminate coupling, but transporting or building a ground screen for a large array on the Moon presents yet another hurdle. Thus, to reiterate, achieving the required levels of foreground subtraction will necessitate accurately modeling the primary beam of one's array, likely at each antenna.

As shown by other 21~cm experiments, it is quite likely that a combination of different in-situ measurement and Bayesian-based statistical techniques will be needed to help constrain the instrument systematics. For example, the antenna beam can be characterized through direct measurement of the lunar regolith properties at the deployment site with ground penetrating radar \citep[e.g.,][]{li2020moon} and impedance measurements \citep[e.g.,][]{altamirano2025using,hendricksen2026estimating}; or characterized through the use of an in-orbit transmitting reference radio signal to probe beam response when an artificial satellite is orbiting pass the antenna main beam \citep[e.g.,][]{neben2015measuring, neben2016hydrogen, bale2023lusee}.  

Another potential alternative is to utilize a semi-realistic beam emulator, such as \texttt{MEDEA} \citep{hibbard2024medea}, which spatially decomposes a set of CST beam models from a range of perturbed hyper-parameters (like regolith dielectric properties) into coefficients on a complete and linear basis. In return, far-field beams for an arbitrary regolith parameter can be generated rapidly from interpolating these coefficients without the need of running more time-consuming CST simulations.

While our analysis is limited to the delay transform power spectrum estimator, the general conclusions drawn from this analysis carry over to image-based power spectrum pipelines. An advantage of image-based estimators is that they Fourier transform along the line-of-sight wavenumber $\eta$ meaning that the transform does not drift over multiple spatial $k_{\perp}$ modes like a delay transform. In theory, given a regularly sampled, dense, and infinite $uv$-plane, foregrounds would be contained to low $k_{\parallel}$ and no wedge would be formed. While it is possible to achieve high levels of wedge suppression by replicating these conditions \citep[see][]{mackay2025completesamplinguvplane}, a full removal of the foreground wedge is likely not possible with the levels of precision required for Dark Ages observation. Additionally, subtle variations per baseline of the electromagnetic coupling between antennas and the ground or other antennas will introduce spectral structure that will inevitably cause a wedge to form.

Therefore, we conclude that it is necessary to acquire high-precision, high-complexity models of the antenna primary beam at Dark Ages frequencies, including but not limited to a detailed model of the lunar regolith and its electromagnetic properties, along with future in-situ measurements at the deployment site. Understanding these phenomena is paramount to observing the small $k_{\parallel}$ modes where the SNR is greatest.

\begin{acknowledgments}

WS and JP acknowledge support for this work from NASA grants 80NSSC21K0693 and 80NSSC22K1745. WS also acknowledges support from a NASA Rhode Island Space Grant Fellowship and from the Brown University Department of Physics. BN is supported by the NSF grants SII NRDZ (AST-2232159) and SWIFT-SAT (AST-2332422). JB was funded by NASA grant 80NSSC22K1264 to support radio astrophysics from the Moon, NASA APRA grant award 80NSSC23K0013, and a subcontract from UC Berkeley (NASA award 80MSFC23CA015) to the University of Colorado (subcontract \#00011385) for science investigations involving the LuSEE-Night lunar far side mission. This work utilized the Blanca condo computing resource, in conjunction to the CST software license funded by JB, at the University of Colorado Boulder. Blanca is jointly funded by computing users and the University of Colorado Boulder. The National Radio Astronomy Observatory (NRAO) and the Green Bank Observatory (GBO) are facilities of the NSF operated under cooperative agreement by Associated Universities, Inc.

\end{acknowledgments}

\vspace{5mm}

\software{21cmFast \citep{10.1111/j.1365-2966.2010.17731.x, Murray2020}, astropy \citep{astropy:2013, astropy:2018, astropy:2022}, lunarsky (\url{ https://github.com/aelanman/lunarsky}), pyuvdata \citep{Hazelton2017, 2025JOSS...10.7482K}, puuvsim \citep{2019JOSS....4.1234L}}


\bibliography{references}{}
\bibliographystyle{aasjournal}

\end{document}